\documentclass[onecolumn,showpacs,amsmath,amssymb,aps]{revtex4}

\usepackage{graphicx}
\usepackage{dcolumn}
\usepackage{bm}
\usepackage{epsfig}

\begin{document}

%%%%%%%%%%%%%%%%%%%%%%%%%%%%%%%%%%%%%%%%%%%%%%%%%%%%%%%%%%%%%%%%%%%%

\markboth{Park, J., Celma, O., Koppenberger, M., Cano, P. and Buld\'u, J.M.}
{Social Network of Contemporary Popular Musicians}

\title{The Social Network of Contemporary Popular Musicians}

\author{Juyong Park$^1$, Oscar Celma$^2$, Markus Koppenberger$^2$, Pedro Cano$^2$ 
and Javier M. Buld\'u$^3$}

\address{
$^1$ Department of Physics and the Center for the Study of Complex Systems, The University of Michigan, Ann Arbor, MI 48109, U.S.A.\\
$^2$Music Technology Group, Universitat Pompeu Fabra, 08003, Barcelona, Spain\\
$^3$Nonlinear Dynamics and Chaos Group,
Departamento de Matem\'aticas y F\'{\i}sica Aplicadas y
Ciencias de la Naturaleza, Universidad Rey Juan Carlos, Tulip\'an s/n,
28933 M\'ostoles, Madrid, Spain.\\
}

\begin{abstract}

In this paper we analyze two social network datasets of contemporary
musicians constructed from \emph{allmusic.com} (AMG), a music and artists' information database: one is the \emph{collaboration network} in which two musicians are connected if they have performed in or produced an album together, and the other is the \emph{similarity network} in which they are connected if they where musically similar according to music experts.  We find that, while both networks exhibit typical features of social networks such as high transitivity, several key network features, such as degree as well as betweenness distributions suggest fundamental differences in music collaborations and music similarity networks are created.
\end{abstract}

\maketitle

%%%%%%%%%%%%%%%%%%%%%%%%%%%%%%%%%%%%%%%%%%%%%%%%%%%%%%%%%%%%%%%%%%%%%%%%%%%%%%%%%%%%%

\section{Introduction}

Developments in computer and information technologies have allowed users to search for information they need on the Internet, rather than in the traditional arena of libraries or printed media.  Particularly, developments in e-commerce technology have produced large commercial retailers of thousand of products serving millions of customers each day.  E-comerce technology has reduced the cost of inventory storage and distribution, leading to what is known as the long-tail phenomenon \cite{and04}. This is related to the distribution of sales of a general item (books, CDs, DVDs, etc.), which generally decays with a power law distribution ---a few items are sold in high volumes while most items suffer low sales volume.  Therefore, by reducing the storage and distribution costs, it can be profitable for companies to concentrate on selling those less popular items, whose total amount can then overcome the incomes of a {\em best-seller}.  Several websites, such as Amazon [http://www.amazon.com], allow on-line users to access any product by navigating through a network of links between items.  Besides the commercial impact of allowing low-sales items to gain visibility, this kind of networks are sources of information about product similarity, category structures, and so forth. 

In this paper, we analyze the topology of two social networks of contemporary popular musicians taken from the AllMusic database of music metadata [http://www.allmusic.com]. The content on the database is created by professional data entry staff, editors and writers. We work with two datasets, that of the \emph{collaboration network} and the \emph{similarity network} of artists in the database.  The networks were constructed as follows: two artists were connected in the collaboration network when they have worked on one or more albums together, while they were connected in the similarity network by music experts of AllMusic according to some criteria.

There are several reasons that make these networks interesting.  First, studying the collaboration network, formed naturally by the actual professional acts of artists, may teach us how musical tendencies spread via formation of profession relationships between musicians, which could prove worthwhile for musicology.  Second, studying the similarity network, which is a large-scale result of human experts' perception of music, may help in inventing recommendation systems in which machines are trained to perform the same
task, and eventually help users discover music easier~\cite{ell02}.

We will see, both networks show typical characteristics of real-world networks such as high transitivity and the small-world property, as some other have shown~\cite{lim03,gle03,can06}.  However, the discrepancies of the two sets, notably in the degrees and the betweenness centralities of same vertices, suggest a fundamental difference between the two networks.

%%%%%%%%%%%%%%%%%%%%%%%%%%%%%%%%%%%%%%%%%%%%%%%%%%%%%%%%%%%%%%%%%%%%%%%%%%%

\section{The Datasets}

On a typical artist's page on the \emph{allmusic.com} database, we can find hyperlinks to other artists under various categories: ``Similar Artists'', ``Worked With'', ``Followers'', etc.  We can regard the existence of a link between two artists as having a tie in a social network.  Using links in the category ``Similar Artists'' that had been created by the music experts of \emph{allmusic.com}, we constructed the \emph{similarity network}. Here, Mick Jagger of the Rolling Stones is connected to Tina Turner or David Bowie.  On the other hand, using links in another category ``Worked With'', we constructed the \emph{collaboration network}, where Mick Jagger is now connected to other members of the Rolling Stones, such as Keith Richards or Charlie Watts, and others~\footnote{Actually the hyperlinks are not reciprocated in this database. The reason for that is the limit space in the HTML page for each artist entry. However, we will treat being similar and having collaborated as mutual. Therefore, we have considered all such links as undirected in the remainder of this work.}.

\begin{figure}[htb]
\includegraphics[width=10cm]{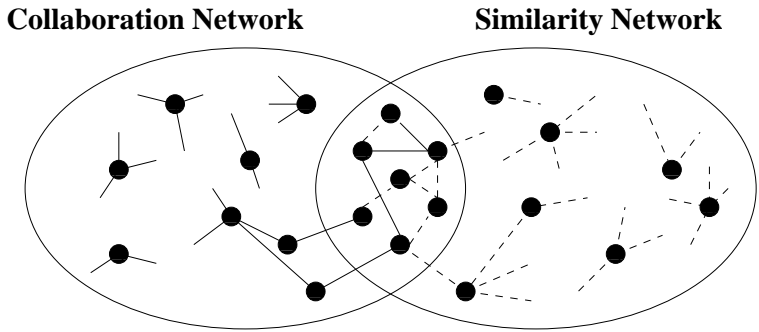}
\caption{The structure of the data sets studied in this paper. Two sets of
network data have an intersection consisting of common vertices.  These common vertices and the edges between them (similarity or collaboration) comprise the subnetworks.}
\label{datasets}
\end{figure}

The similarity network is composed of $32,377$ vertices (artists) and $117,621$ edges, and the collaboration network is composed of $34,724$ vertices and $123,082$ edges.  These two networks have $8,509$ vertices in common.  These common vertices have $24,950$ edges in the similarity network, and $20,232$ edges for the collaboration network, between themselves.  We can visualize this as Fig.~\ref{datasets}.  The two subnetworks defined on these common vertices will enable us to conduct a direct comparison study between similarity and collaboration link patterns.

%%%%%%%%%%%%%%%%%%%%%%%%%%%%%%%%%%%%%%%%%%%%%%%%%%%%%%%%%%%%%%%%%%%%%%%%%%%

\section{Basic Network Properties}

\begin{figure}[t]
\includegraphics[width=125mm]{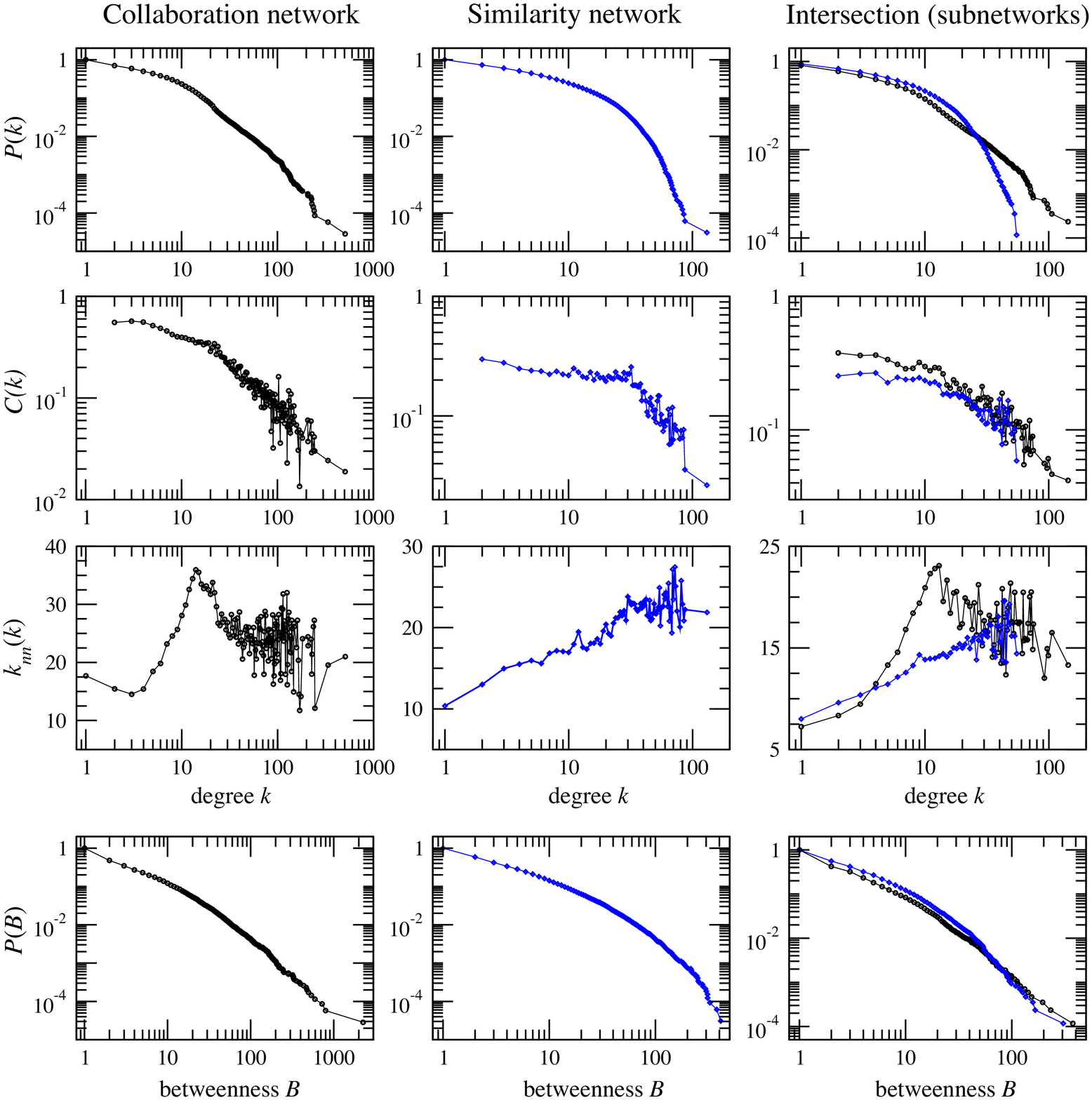}
\caption{The cumulative degree distribution $P(k)$ (first row), the local clustering coefficient $C(k)$ (second row), the nearest--neighbor degree distribution $k_{nn}$ (third row), and the cumulative betweenness centrality distribution $P(B)$ for the collaboration (circle) and similarity (diamond) networks.}
\label{properties}
\end{figure}

In this section, we study several key properties of the networks, such as the degree distribution, transitivity, nearest-neighbor degree correlation, component structure and the Freeman centrality of vertices.  They are summarized in Table~\ref{tab1} and Fig.~\ref{properties}.

\begin{table}[t]
\begin{center}
\begin{tabular}{|c|c|c|c|c|}\hline
%\multirow{2}*{Types}
%&\multicolumn{4}{|c|}{DATA SET}  \\ \hline
& \multicolumn{2}{|c|}{Similarity network} & \multicolumn{2}{|c|}{Collaboration network}\\ \hline
& entire & intersection & entire & intersection \\
\hline %\hline
%%%%%%%%%%%%%%%%%%%%%%%%%%%%%%%%%%
$n$           &  32\,377  &   8\,509  &   34\,724  & 8\,509 \\
$m$           & 117\,621  &  24\,950  &  123\,122  & 20\,232 \\ \hline
size of $S_0$ & 30\,384~(94$\%$)  &  7\,219~(85$\%$)  & 30\,945~(89$\%$) & 6\,054~(71$\%$) \\ 
$\bar{d}~(d_{\mathrm{max}})$ & 6.5~(22)  &  6.0~(20)  &  6.4~(23)  & 6.3~(19) \\ \hline 
$C$           & 0.185~($18.5\%$) &  0.178~($17.8\%$)  &  0.182~($18.2\%$) & 0.171~($17.1\%$) \\ \hline
              &  $131$ & $55$ & $508$ &  $143$ \\
$k_{\mathrm{max}}$  &R.E.M.&Eric Clapton&P. Da Costa&P. Da Costa\\
              & & & &R. Van Gelder\\
\hline
highest-betweenness & Sting & Sting & P. Da Costa & P. Da Costa \\
artist & & & & \\
%%%%%%%%%%%%%%%%%%%%%%%%%%%%%%%%%%
% The Gamma of Art of the Mix is 1.45 which, according to Dorogotvsev
% corresponds to a network whose edges are growing faster than its
% vertices.
%\hline
\hline
\end{tabular}
\small
\caption{
Summary of several network characteristics of similarity, collaboration, and intersection subnetworks: number of vertices $n$, number of edges $m$, number of vertices in the largest component $S_0$ and its percentage among all vertices, mean geodesic path $\bar{d}$ in $S_0$, diameter $d_{max}$ of $S_0$, global clustering coefficient $C$, the highest-degree $k_{\max{max}}$ and the corresponding artist(s), and the artist with the highest betweenness.
}
\label{tab1}
\end{center}
\end{table}
\normalsize

%\begin{table}
%\begin{tabular}{|c|c|c|c|c|c|c|c|c|c}\hline
%  \multicolumn{2}{|c|}{Data set} & $n$ & $m$ & Size of the &
%  $\bar{d}(d_{\mathrm{max}})$ & $C$ & $r$ &
%  $k_{\mathrm{max}}$ (Artist) & Artist with \\
%  \multicolumn{2}{|c|}{}   &     &     & largest component &     &     &
%   & max. $B$\\ \hline
%collaboration & entire & $34724$ & $123122$ & $30945$ ($89\%$) & & $0.182$
%  & $-0.006(2)$ & $508$ (P. Da Costa) & P. Da Costa \\ \cline{2-9}
%              & intersection & $8509$ & $20232$ & $6054$ ($71\%$) &
%  $6.3 (19)$ &
%  $0.171$ &
%  $0.037(5)$ & $143$ (R. Van Gelder / P. Da Costa) & P. Da Costa \\ \hline
%similarity & entire & $32377$ & $117621$ & $30384$ ($94\%$) & & $0.185$
%  & $0.185(3)$ & $131$ (R.E.M.) & Sting \\ \cline{2-9}
%              & intersection & $8509$ & $24950$ & $7219$ ($85\%$) &  &
%  $0.178$ &
% $0.188(6)$ & $55$ (Eric Clapton) & Sting \\ \hline
%\end{tabular}
%\caption{Some basic network properties of the data sets.}
%\label{tab1}
%\end{table}

%%%%%%%%%%%%%%%%%%%%%%%%%%%%%%%%%%%%%%%%%%%%%%%%%%%%%%%%%%%%%%%%%%%%%%%%%%%

\subsection{Mean geodesic length, diameter and component structure}

A prominent feature of a complex network is the called the ``small-world effect'' \cite{str98} which means that the shortest paths (also called geodesics) between vertices is very small compared to the system size.  The longest geodesic in the network is called its diameter.  We see in Table~\ref{tab1} that average geodesic length $\bar{d}$ is smaller than $7$, while the diameter is no larger than $23$ in each network.

A component of a network is the set of vertices that are connected via one or more geodesics, and disconnected (i.e., no geodesics) from all other vertices. Typically, networks possess one large component that contains a majority of vertices. In Table~\ref{tab1} we see that with the exception of the collaboration subnetwork, each giant component contains $\sim 90\%$ of the vertices.

\subsection{Degree distribution}

%Another interesting feature is the degree distribution $p(k)$.
The number of vertices linked to a vertex is called its degree, usually denoted $k$.  The degree distribution $p_k$ is the fraction of vertices in the system with degree $k$.  Many real-world networks, including the Internet and the worldwide web (WWW), are known to show a
right-skewed distribution, often a power law $p_k\propto k^{-\tau}$ with $2<\tau<3$.  More frequently, the cumulative degree distribution $P(k) = \sum_{k'=k}^{\infty}p_{k'}$, the fraction of vertices having degree $k$ or larger, is plotted.  A cumulative plot avoids fluctuations at the tail of the distribution and facilitates the evaluation of the power coefficient $\tau$ in case the network follows a power law. 

We see in Figure~\ref{properties} that collaboration network exhibit power-law degree distributions near their tails, $p(k)\sim k^{-3}$, following a straight line in a log-log representation.  We obtain a similar result when looking at its intersection subnetwork.  
%As a curiosity, in the entire data set, the Brazilian percussionist Paulinho Da Costa has the largest degree of $508$ (i.e. he has worked with $508$ different artists) and in the  intersection set he is tied with the legendary  recording engineer Rudy Van Gelder. {\bf (What are you trying to say? I don't get it!)}

On the other hand, the similarity network closely follows an exponential form of $p_k\sim \exp{-0.12k}$, while its intersection subnetwork follows $p_k\sim \exp{-0.15k}$.  As such, there is a huge difference $k_{\mathrm{max}}$: artist R.E.M. and Eric Clapton are the most connected in the entire dataset and the subnetwork with degrees $131$ and $55$ respective, while in the collaboration dataset, Paulinho Da Costa tops in both cases with degrees $508$ and $143$ (tied with the legendary recording engineer Rudy Van Gelder of the famed Blue Note label), several times larger than his counterparts in the similarity network.

%%%%%%%%%%%%%%%%%%%%%%%%%%%%%%%%%%%%%%%%%%%%%%%%%%%%%%%%%%%%%%%%%%%%%%%%%%%

\subsection{Transitivity}
Transitivity, or clustering, is an indication of how cliquish (tightly knit) a network is. It is quantified by the abundance of triangles in a network, where a triangle is formed when three vertices are all linked to one another. It can be quantified by the 
\emph{global clustering coefficient} $C$, defined as 
\begin{equation}
C = \frac{3\times\textrm{number of triangles}}{\textrm{number of connected triples}}.
\end{equation}
Here, a connected triple means a pair of vertices connected via another vertex.  Since a triangle contains three triples, $C$ is equal to the probability that two neighbors of a vertex is connected as well.  Typical social networks have $C$ of fractions of percent, and in Table~\ref{tab1} that the music networks show values of $17\%$--$18\%$.  The reason that this indicates an abundance of triangles is that in the most random graph model of comparable size ($n$ and $m$), $C$ is almost negligible --- for example, with $n=34\,724$ and $m=123\,082$, a random graph has $C=0.02\%$.

A closely related yet distinct measure is the \emph{local clustering coefficient} $C_i$ of each vertex $i$ (defined for the case $k_i>1$) defined as
\begin{equation}
C_i = \frac{\textrm{number of connected pairs of neighbors of
    $i$}}{\textrm{number of pairs of neighbors of $i$}=\frac{1}{2} k_i(k_i-1)},
\end{equation}
which is the fraction of pairs of neighbors of a vertex are connected.

Often the local clustering is plotted as a function of degree $k$ defined as the average of $C_i$ over all vertices with a given degree $k$:
\begin{equation}
C(k)=\langle C_i \rangle \bigl|_{k_i=k}.
\end{equation}

Some real-world networks are known to show a behavior of $C(k)\propto k^{-1}$, usually attributed to the hierarchical nature of networks~\cite{rav02}.  In Fig.~\ref{properties} we have plotted the local $C(k)$. We observe that $C(k)$ decreases as $k^{-1}$ for range $30\lesssim k\lesssim 300$ for the collaboration network, but the decreasing pattern is not as clear for other data sets.
%This result reveals the coexistence of a hierarchy of vertices (artists) with different degrees of modularities (i.e., artists belong to gangs of ``close" artists which have different sizes), since the local clustering coefficient (which measures in a certain way the modularity of the network) scales with a power law with the number of neighbors $k_i$ (leading to a hierarchy).

%%%%%%%%%%%%%%%%%%%%%%%%%%%%%%%%%%%%%%%%%%%%%%%%%%%%%%%%%%%%%%%%%%%%%%%%%%%

\subsection{Degree correlations}

We have also calculated the average nearest-neighbor degree
$k^{\mathrm{nn}}$ as a function of $k$,
\begin{eqnarray}
k^{\mathrm{nn}}(k) = \sum_{k'=0}^{\infty}k'p(k'|k),
\end{eqnarray}
where $p(k'|k)$ is the fraction of edges that are attached to a vertex
of degree $k$ whose other ends are attached to vertex of degree
$k'$. Thus $k^{\mathrm{nn}}$ is the mean degree of the vertex we find by
following a link emanating from a vertex of degree $k$.

The $k^{\mathrm{nn}}$ for our four datasets are plotted in
Fig.~\ref{properties} (third row). Here we see another difference between the two
main networks [apart from that observed in $P(k)$].  While for the similarity network it is a nearly monotonic,
increasing function, for the collaboration network it is not at all a
simple form.
The evolution of $k_{nn}(k)$ is related with the {\em assortativity} of the 
network \cite{new02}, which indicates the tendency of a vertex of degree $k$
to associate with a vertex of the same $k$. When $k_{nn}(k)$ is an increasing
function of $k$, which is the case of the similarity network 
(see Fig.~\ref{properties}, third row, central plot), the network is
assortative. In other words, the most connected artists are prone to be
similar to other top connected artists. On the other hand, we can observe that the
collaboration network is rather noisy 
(Fig.~\ref{properties}, third row, first plot). The first section of the $k_{nn}(k)$, for values up to 12 is assortative while the tail is not. The assortativeness for small values of $k$ could relate with band size in which all components obviously collaborate with all the others. The same reasoning does not apply for larger values. It could also be argued the assortativity observed in the similarity network is not a consequence
of collaboration between artists. 

  A closely-related concept is the {\em degree-degree
correlation coefficient} $r$, which is the Pearson correlation
coefficient for degrees of vertices at either end of a link:

\begin{eqnarray}
r &=& \frac{\sum_ik_i^2k_i^{\mathrm{nn}} -
     (2m)^{-1}\bigl[\sum_ik_i^2\bigr]^2} {\sum_ik_i^3 -
     (2m)^{-1}\bigl[\sum_ik_i^2\bigr]^2} \simeq
      \frac{\sum_kk^2k^{\mathrm{nn}}(k)p_k -
     z^{-1}\bigl[\sum_kk^2p_k\bigr]^2} {\sum_kk^3p_k -
     z^{-1}\bigl[\sum_kk^2p_k\bigr]^2},
\end{eqnarray}
where $p_k$ is the degree distribution, $\sum_i$ denotes sum over
vertices and $\sum_k$ denotes sum over degrees. We can clearly see the
connection between $r$ and $k^{\mathrm{nn}}(k)$. In the case of a
monotonically increasing (decreasing) $k^{\mathrm{nn}}(k)$ which means,
as mentioned before,
that high-degree vertices are connected to other high-degree
(low-degree) vertices and vice versa, it results in a positive value
of $r$, as in the case of the similarity network which has $r=0.184$
(for the intersection portion of the network, $r=0.188$).  In other
cases, however, we cannot read it off easily: for the entire collaboration
network, $r=-0.00575$ while for its intersection portion $r=0.0372$.
The collaborative networks result, especially, is an interesting observation, since most social
networks are known to show positive degree-degree correlation (as seen in the 
similarity network), and it
is thought to be originating in part from the community structure \cite{new03}.

%%%%%%%%%%%%%%%%%%%%%%%%%%%%%%%%%%%%%%%%%%%%%%%%%%%%%%%%%%%%%%%%%%%%%%%%%%%

\subsection{The betweenness (Freeman) centrality}

Given the inhomogeneity of link patterns around vertices in a
complex network, we could certainly imagine that the position and
roles of vertices will vary significantly from one vertex to
another. {\em Centrality}, as its name suggests, is a concept that differentiates vertices
according to how influential, or \emph{central}, they are in a
network.  Degree is one kind of centrality, since it would be
reasonable to assume that people with particularly many acquaintances
can be looked as being important figures.  However, degree is
primarily local in scope (and talking loudly does not mean you are
affecting others more effectively than somebody who speaks quietly but
very eloquently, so to speak), and to overcome its shortcomings social
scientists have in particular developed various measures of
centrality.  For our networks' dataset we choose to study the
{\em betweenness} or {\em Freeman} centrality \cite{fre77}.

The idea behind this centrality measure is that a central vertex will
act as a relay of information between vertices, a role endowed
thanks to being on a geodesic between vertices (hence the name
betweenness).  Considering a vertex has a relay of information so
that it has a ``power to withhold information \ldots  or to refuse to
pass on requests for information'' seems intuitively
appropriate for communication networks systems, and recently has been
studied on the Internet as well \cite{goh02,vaz02}.

The reason for choosing this centrality to study these networks
was that we were interested in gaining a glimpse of how musical
influences (considered as information) might spread via the complex network of artists.
Especially, ``crossover'' musicians are becoming more common these
days, and we were anticipating that those people who produce albums
across genres were important in musical developments of the multiple
genres, and by becoming bridges between genres, might have higher
betweenness centrality.

The definition of {\em Freeman (betweenness) centrality} $B_l$ of a
vertex $l$ is defined as
\begin{eqnarray}
B_l = \frac{1}{2}\sum_{i,j}\frac{g_{ilj}}{g_{ij}},
\end{eqnarray}
where $g_{ij}$ is the total number of geodesics between vertices $i$ and
$j$, and $g_{ilj}$ is the number of the ones that pass through
the vertex $l$.

In Fig.~\ref{properties} (fourth row) we have plotted the \emph{cumulative} 
fraction $P_B(k)$ of Freeman
centralities for our datasets. We see that this
distribution is highly skewed for both cases (with no differences at
the subnetworks). Similar results were obtained 
by different authors \cite{goh02,vaz02} in other kind of networks. 
In Table~\ref{tab1} there is list of
artists with the highest betweenness centrality in each data set. It
is interesting to note that in the cases of similarity network data,
the highest-degree vertex is not the highest-centrality vertex. We will
discuss this point deeply in the next section. 

%%%%%%%%%%%%%%%%%%%%%%%%%%%%%%%%%%%%%%%%%%%%%%%%%%%%%%%%%%%%%%%%%%%%%%%%%%%

%%%%%%%%%%%%%%%%%%%%%%%%%%%%%%%%%%%%%%%%%%%%%%%%%%%%%%%%%%%%%%%%%%%%%%%%%%%

\section{Comparison of self-organized network and artificial network}\label{discuss}

An interesting question, as we have posed in the beginning of this
paper, is how differently an individual is represented in different
types of networks.  People belong to many spheres of social activity,
and their relationship with the same people may well be different in
each sphere. In fact, the two of our intersection data set seem to be quite
different. Among the $24,950$ and $20,232$ edges belonging to the
collaboration and similarity data respectively, there are only $464$
common edges,
so having worked together does not necessarily
(practically not at all) translate into being classified as musically
similar.

To see how different the individuals' roles are in these two
networks, in Table~\ref{rank} we have indicated the top ten high
betweenness scorers from either network, along with their ranks in the
other data set. The difference is evident. For example, Paulinho Da
Costa, the prolific Brazilian percussionist, ranked first in the
collaboration network, is ranked at merely
$2,933^{\mathrm{th}}$ in the similarity network.  On the other hand,
Sting, ranked at the top in the similarity network, is ranked at
$1,406^{\mathrm{th}}$ in the collaboration network.  In fact, none of
the top ten artists in either network is ranked as high in the other
network. Quantitatively, the Spearman correlation of the two ranks is
$0.255$, indicating that the two are only slightly correlated.
%(a Pearson correlation for the betweenness per se is $0.203$), therefore

%%%%%%%%%%%%%%%%%%%%%%%%%%%%%%%%%%%%%%%%%%%%%%%%%%%%%%%%%%%%%%%%%%%%%%%%%%%

\begin{table}
\begin{tabular}{|c|c|c|c|}  \hline
\multicolumn{4}{|c|}{COLLABORATION} \\ \hline
Rank & Artist            & rank in             & comments \\
     &                   & similarity network &  \\ \hline
1    & Paulinho Da Costa & 2,933                & Percussionist \\
2    & Jim Keltner       & 5,468                & Percussionist \\ 
3    & Ron Carter        & 2,689                & Bassist\\
4    & Rudy Van Gelder   & 5,468                & Recording engineer\\
5    & Dean Parks        & 5,468                & Guitarist\\
6    & Herbie Hancock    & 299                 & Jazz pianist\\
7    & Randy Brecker     & 4,073               & trumpetist and flugenhornist\\
8    & Jim Horn          & 4,517                & Saxophonist\\
9    & Dann Huff         & 3,620                & Guitarist \\
10   & Tony Levin        & 1,471                & Bassist\\ \hline \hline
\multicolumn{4}{|c|}{SIMILARITY} \\ \hline
Rank & Artist            & rank in               & comments \\
     &                   & collaboration network &  \\ \hline
1    & Sting             & 1,406                  & singer, bassist\\
2    & Joni Mitchell     & 837                   & singer, song writer\\
3    & Eric Clapton      & 23                    & guitarist, singer\\
4    & Quincy Jones      & 46                    & producer, trumpeter\\
5    & Gil Evans         & 396                   & jazz pianist\\
6    & Jimi Hendrix      & 3,047                  & guitarist, singer\\
7    & M. Davis$^1$/C. Parker$^2$ & 41           & trumpeter$^1$, saxophonist$^2$\\
8    & Aretha Franklin   & 67                    & singer\\
9    & Lenny Kravitz     & 2,446                  & singer, songwriter\\
10   & Jeff Beck         & 463                   & guitarist\\ \hline
\end{tabular}
\caption{The ten top-ranked artists in betweenness in either of the
  intersection dataset, with their ranks in the other data set
  indicated. The two ranks are moderately
  correlated with Spearman coefficient $0.255$. }
\label{rank}
\end{table}
 
If we look at Table~\ref{rank} in more detail,
we can see each vertex's characteristics and/or specialty in action.  Artists
with the largest betweenness in the collaboration data set are
primarily instrumentalists (except 
for Rudy Van Gelder, a prolific recording engineer of the famed Blue
Note and Verve labels, among many), and indeed all nine musicians are
most famous for their virtuosity in the indicated instruments.
They must have been invited to work in a multitude of
recording sessions for various projects (in our data set, Paulinho da
Costa has had $143$ collaborators in the 
intersection data set, and $508$ overall in the entire
data set of collaborations), possibly bridging musicians of
different styles to result in a high betweenness. However that did not
necessarily translate into their perceived musical styles becoming as varied.
A possible explanation for that is that some musicians adapt to the style of music
that the recording artists requires.

Considering the similarity network, it is remarkable to find
an exponential decay in their degree distribution, since many
social networks exhibit a power law \cite{new03}. Nevertheless
we must be very cautious since the similarity network has been 
designed by human perception (the opinion of experts). In this way,
the evaluation of how similarity (i.e. musical tendencies) spreads will
always be filtered by a subjective opinion, a fact that may cover (and filter)
the real structure of the similarity network. In this sense, efforts
have been made during the last years in order to obtain numerical
algorithms to evaluate, in a rigorous and objective way, similarity
between songs (and artists) \cite{ell02}. Nevertheless, how to capture music similarity
as perceived by humans is still an open field.

\section{Conclusions}

In this paper, we have looked at various network properties of two
types of music networks. One was the collaboration relations among
musicians which must have evolved naturally, and
the other was the musical similarity amongst them, which was entirely
constructed via human perception of music. 
We have analyzed the structural properties of the networks,
observing that both networks
share small world properties together with a clustering coefficient
following a power law. The latter indicates the existence of a certain
modularity that depends on the vertex degree (leading to a hierarchy). In this
way, better connected artists form larger clusters than those of artists
with less connections.  
Despite networks are constructed with artists
as vertices and a certain connection between them (similarity/collaboration),
we obtain different results, such as the degree distribution, which follows a power
law in the collaboration network and has exponential decay in the similarity
network. In addition, the Freeman centrality shows that vertices with highest
betweenness are completely different at both networks, a fact that indicates
that collaboration is not the mechanism for similarity spreading. Reciprocally, playing similar music is not an ingredient to predict collaboration links. It is indeed usual that artists collaborate with artists from a complete different style. 
The difference between the similarity and collaboration networks rules out the possibility of using collaboration data to infer music similarity. This would have proved convenient because collaboration data is easier to gather and definitely more objective.

%%%%%%%%%%%%%%%%%%%%%%%%%%%%%%%%%%%%%%%%%%%%%%%%%%%%%%%%%%%%%%%%%%%%%%%%%%%

\end{document}